\documentclass{elsart3}
\usepackage{graphics}
\usepackage{epsfig}
\usepackage{amssymb,amsmath}
\begin{document}
\begin{frontmatter}

\title{Localised magnetic excitations of coupled impurities in a
 transverse
Ising ferromagnet}

\author{R. V. Leite,}
\ead{valmir@fisica.ufc.br}
\author{J. Milton Pereira Jr.}
\ead{pereira@fisica.ufc.br}
\author{ and R. N. Costa Filho}
\ead{rai@fisica.ufc.br}

\address{Departamento de F\'{\i}sica, Universidade Federal do
Cear\'a, Caixa Postal 6030\\ Campus do Pici, $60451-970$
Fortaleza, Cear\'a, Brazil}

\date{\today}

\begin{abstract}
A Green's function formalism is used to calculate the spectrum of
excitations of two neighboring impurities implanted in a semi-infinite
ferromagnetic. The equations of motion for the Green's functions are
determined in the  framework of the Ising model in a transverse field
and results are given for the  effect of the exchange coupling,
position and orientation of the impurities  on the spectra of
localized spin wave modes.
\end{abstract}

\begin{keyword}
Heisenberg ferromagnet; Ising model-transverse;  Green's function;
Spin wave; Impurities modes

\PACS 75.30.Hx  \sep 73.30.Ds \sep 75.30.Pd
\end{keyword}
\end{frontmatter}

\section{Introduction}
Solids are known to support a variety of elementary excitations, such
as phonons, polaritons and magnons. In order to model the behavior of
these excitations, one often makes the assumption that the media in
which they propagate can be treated as a perfect crystal with infinite
extension. This approximation, however, does not alway give a reliable
picture of the system. An example is given by low-dimension media,
such as ultra-thin films, in which the presence of surfaces has been
shown to strongly influence their spectra of excitations. That is also
true for media containing impurities or defects. These different
behaviors occur due to the fact that the presence of interfaces or
defects in an otherwise ideal medium can modify the microscopic
interactions in the material. Also, the low dimensionality of the
medium, together with the presence of  impurities, acts to break the
translational symmetry of the system, causing significant
modifications on the propagation of the excitations, and also allowing
the existence of localized excitation modes.

In the case of magnetic materials, the properties of localized
excitations such as surface and impurity modes have been widely
studied, theoretically as well as experimentally
\cite{MikeTilley,Chen1,Chen2,Henkel99} and several models have been
proposed to elucidate the dynamics of these modes. Among these, the
transverse Ising model, in particular, has been shown to give a good
theoretical description of real materials with anisotropic exchange
(e.g., CoCs$_{3}$Cl$_{5}$ and DyPO$_{4}$) and of materials in
which the crystal field ground state is a singlet \cite{Wong}.

The Ising model in a transverse field has been applied
to ferromagnets to obtain the spin wave (SW) spectrum of semi-infinite
media \cite{Tilley,Shiwai} and films \cite{Salman,Kontos,Blinc}. The
SW frequencies associated  with the presence of a impurity layer in an
otherwise uniform semi-infinite medium has also attracted attention.
In both cases the impurities, while breaking the translational
symmetry along the  direction perpendicular to the surface of the
media, were assumed to be uniformly distributed  along the plane
of the film layers. The results showed the presence of localized
impurity modes which were found to depend on the position of the
impurity layer within the  film, as well as on the strength of the
exchange coupling between the magnetic impurity  sites.
A different aspect of this problem is to consider the effect of the
presence of localized impurities in the medium. In the present paper
we make an extension of the theory presented in  Ref.\cite{Costa}, by
considering the effect of several localized impurities in an otherwise
pure semi-infinite ferromagnet, in the context of the transverse Ising
model. Numerical results are obtained by means of a Green's function
technique. The paper is structured as follows: in section II the
theoretical method is introduced. Numerical results are presented and
discussed in section III. In section IV the main results are
summarized and conclusions are presented.

\section{Model and Green's function formalism}

We consider a semi-infinite ferromagnet with a (001) surface and a
simple cubic structure (lattice constant $a$). Two nearest- neighbor
 localized
impurities spins are taken to be embedded in the medium at distance
$(n-1)a$ from the surface (where integer $n\geq1$). The localized
spins are described by the transverse Ising model and the Hamiltonian
of the system in the presence of an external field can be written as
\begin{equation}
{\mathcal H}= -\frac{1}{2}\sum_{l,m}J_{lm}S_l^zS_m^z-\sum_lh_lS_l^x
\end{equation}
where $S_{l,m}^x$ and $S_{l,m}^z$ are the $x$ and $z$ components of
the spin operator
${\mathbf S}$, with $S=\frac{1}{2}$ for all sites. The first term in
the right hand side of Eq. (1) contains the contribution due to
the exchange interaction. Throughout this paper we assume that the
summation runs over nearest neighbor sites. The second term
on the right hand side of Eq. (1) refers to the effect of the
transverse magnetic field in a given site $l$ (this field is $h$ for
 host sites in
the bulk and $h_S$ for host sites at the surface).  For convenience,
 we shall re-express
the Hamiltonian as ${\mathcal H}={\mathcal H}_0 +{\mathcal H}_I$,
where ${\mathcal H}_0$ is the Hamiltonian of a pure host ferromagnet,
whereas ${\mathcal H}_I$ corresponds to the perturbation caused by
impurities.
\begin{equation}
{\mathcal H}_I= {\mathcal H}_{od}+{\mathcal H}_{o'd'}
+{\mathcal H}_{oo'}
\end{equation}
where the ${\mathcal H}_{od}$ (${\mathcal H}_{o'd'}$) terms
contain the  exchange coupling between an impurity at a given site
labeled $o$ ($o'$) and its neighboring host sites. These terms can be
written as
\begin{equation}
{\mathcal H}_{od}= -(J_o-J)\sum_d S_d^zS_o^z-(h_o-h)S_o^x,
%\nonumber
\end{equation}
\begin{equation}
{\mathcal H}_{o'd'}=
 -(J_{o'}-J)\sum_{d'}S_{d'}^zS_{o'}^z-(h_{o'}-h)S_{o'}^x,
%\nonumber
\end{equation}
\begin{equation}
{\mathcal H}_{oo'}= -(J_{I}-J)S_o^zS_{o'}^z.
%\nonumber
\end{equation}
where the $d$ ($d'$) index labels the nearest neighbor host sites.
The exchange constant assumes the values $J_o=J'$ ($J_{o'}=J''$) for
the interaction between the impurities and host sites in the interior
of the medium and $J_o=J_S'$ ($J_{o'}=J_S''$) when both sites are at
the surface and $J$ otherwise. Likewise, the term ${\mathcal H}_{oo'}$
in the Eq.(2) expresses the  exchange interaction between the
impurities. The Zeeman contribution in Eqs.(3) and (4)
describes the effect of the transverse magnetic field at the
 impurities,  $h_o$
 and $h_{o'}$, which assume the values $h'_S$ and $h''_S$ if those 
 are located 
at the surface of the  medium, and $h'$ and $h''$ otherwise.

\begin{center}
\begin{figure}
\includegraphics[width=1.0\linewidth]{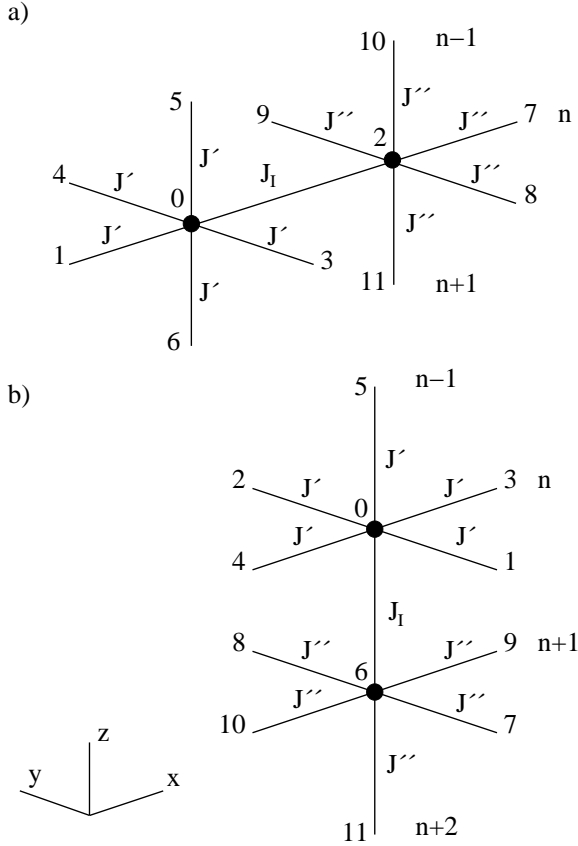}
\caption{Representation of the interaction scheme for two nearest
neighbor impurities in a Ising ferromagnet in a transverse field. The
impurities (black circles) are exchange coupled with each other and
with their nearest neighbors in the host medium, and may be aligned
along (a) the $x$ axis or (b) the $z$ axis.}
\end{figure}
\end{center}

Since we are considering only nearest neighbor spin interactions,
we can distinguish two physically distinct orientations for the
impurities, namely: one with the two impurities aligned along the $x$
axis, which we will refer to as the $X$ case (Fig. 1a) and another,
$Z$ case (Fig. 1b) with two impurities aligned along the $z$ axis. In
order to obtain the excitation spectra of these systems, we extend the
formalism for a single localized impurity presented in
Ref.\cite{Costa}. Thus, we start by defining the Green's functions
$\langle \langle S_l^{\alpha};S_m^{\beta}\rangle \rangle_{\omega}$,
where $\alpha$ and $\beta$ stand for the Cartesian components of the
spin operators and $\omega$  is a frequency label. In the present
paper, we use the retarded commutator Green's function
$G_{lm}({\mathbf q}_{\parallel},\omega)= \left\langle \left\langle
S_l^z;S_m^z\right\rangle  \right\rangle_{\omega }$.
These functions must satisfy the equation of motion \cite{Zubarev}
\begin{equation}
\begin{split}
\omega \left\langle \left\langle S_l^z;S_m^z\right\rangle
 \right\rangle
_{\omega }=&\frac{1}{2\pi }\left\langle \left[ S_l^z,S_m^z\right]
\right\rangle \qquad \qquad \\
& \quad +\left\langle \left\langle \left[ S_l^z,{\mathcal H}\right]
;S_m^z\right\rangle \right\rangle _{\omega}.
\end{split}
\end{equation}
Previous calculations for the pure system showed that a second order
phase transition should occur at a temperature $T_C$, with
$\tanh{(h/2k_BT_C)}=h/3J$, where $k_B$ denotes Boltzmann's  constant,
such that for $T<T_C$ the average spin orientation at each site can
have components in the $x$ and $y$ directions, whereas for $T>T_C$ it
lies along the $x$ direction. The presence of impurities in the system
is expected to change the critical temperature to $T_C^i$, which in
some cases may be  greater than $T_C$. In order to simplify the
calculations, in this work we focus on the  high-temperature regime
(i.e. $T>T_C^i$). The Green's function for the pure system can be
obtained by solving  Eq. (6) with ${\mathcal H}$ replaced by
${\mathcal H}_0$. The solution, which describes an ideal semi-infinite
Ising ferromagnet (i.e., with translational symmetry parallel to the
surface), is well known and is given by  \cite{Cot.}
\begin{equation}
G^0_{lm}(\omega)=-\frac{1}{M}\sum_q G^0_{{\mathbf
q}_{\parallel}}(\omega)\exp{[{\mathbf  q}_{\parallel}\cdot ({\mathbf
r}_l-{\mathbf r}_m)]},
\end{equation}
where the vectors ${\mathbf r}_l$ and ${\mathbf r}_m$ indicate the
positions of two given sites $l$ and $m$, ${\mathbf
q}_{\parallel}\equiv(q_x,q_y)$ is an in-plane wave vector and $M$ is
the number of sites in any layer parallel to the surface.
The Fourier amplitudes of this function are found as
\begin{equation}
\begin{split}
G^0_{{\mathbf q}_{\parallel}}(\omega)=&\frac{1}{2\pi J(x-x^{-1})} \\
&\times\left(x^{|n-n'|}-\frac{1+x^{-1}\Delta}{1+x\Delta}x^{n+n'}\right
).
\end{split}
\end{equation}
Here the labels $n$ and $n'$ are layer indices for the sites $l$ and
$m$, respectively (with $n=1$ being the surface layer.), and $x$
is a complex number that satisfies the condition
\begin{equation}
x+x^{-1}=[h^2-4hJR^x\gamma({\mathbf q}_{\parallel})-\omega^2]/hJR^x
\end{equation}
where $\gamma({\mathbf q}_{\parallel})$ is a structure factor that, in
the case of a simple cubic lattice, is given by $\gamma({\mathbf
q}_{\parallel})=\frac{1}{2}[\cos(q_xa)+\cos(q_ya)]$. In addition,
when describing localized modes the condition $|x|\leq 1$ must also be
fulfilled. The parameter $\Delta$ in Eq. (8) contains the information
regarding the surface and is given by
\begin{equation}
\begin{split}
\Delta=&\frac{\omega^2(h_SR^x_S-hR^x)-hh_S(hR^x_S-h_SR^x)}{hh_SR^xR^x_
SJ}\\ & -4\gamma({\mathbf q}_{\parallel})\left(\frac{J_S}{J}-1\right),
\end{split}
\end{equation}
where, from mean field theory, we obtain the spin averages at the bulk
and surface as $R^x\equiv \langle S^x \rangle = \tanh{(h/2k_BT)}$ and
$R_S^x = \tanh{(h_S/2k_BT)}$, respectively.

The presence of localized impurities in an otherwise ideal medium acts
to break the translational invariance of the system. Consequently, the
calculations for the impure system must be performed in real space.
By including the effects of the impurity in the Hamiltonian and
applying Eq. (8), one can obtain a new Green's function
$G_{lm}(\omega)$, which is found to obey the equations
\begin{equation}
\begin{split}
A_{lj}G_{lm}(\omega)=& \quad\delta_{lj}-
\\&\frac{2\pi}{R^x_{l}}[P_{lj}-U_{lj}-U'_{lj}-U''_{lj}]G_{lm}(\omega)
\end{split}
\end{equation}
where
\begin{equation}
A_{lj}=\frac{\omega^2-[h+(h_o-h)\delta_{lo}+(h_{o'}-h)\delta_{lo'}]^2}
{h+[h+(h_o-h)\delta_{lo}+(h_{o'}-h)\delta_{lo'}]}\delta_{lj},
\nonumber
\end{equation}
\begin{equation}
P_{lj}=\sum_pJ_{lp}R_l^x\delta_{lj}
\nonumber
\end{equation}
\begin{equation}
U_{lj}=\sum_{d}(J_o-J)\,R_l^x
[\delta_{lo}\delta_{dm}+\delta_{ld}\delta_{jo}],
\nonumber
\end{equation}
\begin{equation}
U'_{lj}=\sum_{d'}(J_{o'}-J)\,R_l^x
[\delta_{lo'}\delta_{jd'}+\delta_{ld'}\delta_{jo'}],
\nonumber
\end{equation}
\begin{equation}
U''_{lj}=(J_I-J)\,R_l^x
[\delta_{lo'}\delta_{jo}+\delta_{lo}\delta_{jo'}].
\nonumber
\end{equation}
where the index $o$ ($o'$) assumes the values $0$ ($2$) for each
impurity site in the $X$ case and $0$ ($6$) in the $Z$ case (see Fig.
1). The second summation runs over $5$ neighboring host sites to
either  impurity. By rewriting Eq. (11) in matrix form we obtain the
Dyson equation
\begin{equation}
[({\mathbf {\tilde{G}}}^0(\omega))^{-1}-{\mathbf V}]{\mathbf
 {\tilde{G}}(\omega)}={\mathbf I},
\end{equation}
where ${\mathbf {\tilde{G}}}^0(\omega)$ and
${\mathbf{\tilde{G}}}(\omega)$ are square matrices with elements given
by $(2\pi/R_l^x)G^0_{lm}(\omega)$ and $(2\pi/R_l^x)G_{lm}(\omega)$,
respectively. ${\mathbf I}$ is the unit matrix, ${\mathbf V}$ is an
effective potential related to the impurity term ${\mathcal H}_I$,
with elements
\begin{equation}
\begin{split}
V_{lj}=&\frac{\omega^2-h^2}{h}\delta_{lj}-A_{lj}-P_{lj}
\\&-U_{lj}-U'_{lj}-U''_{lj}.
\end{split}
\end{equation}

Thus, Eq. (9) can be said to relate the matrix Green's function
of the pure (${\mathbf G^0}(\omega)$) and impure (${\mathbf
 G}(\omega)$) systems.
The spectrum of localized modes is then found by numerically
calculating the frequencies that satisfy the determinantal condition
\begin{equation}
\det[\mathbf {I}-\mathbf {\tilde{G}}^0(\omega)\mathbf{V}]=0
\end{equation}
which gives the poles of the Green's function for the impure system.

\begin{figure}[]
\resizebox{0.55\textwidth}{!}{\includegraphics{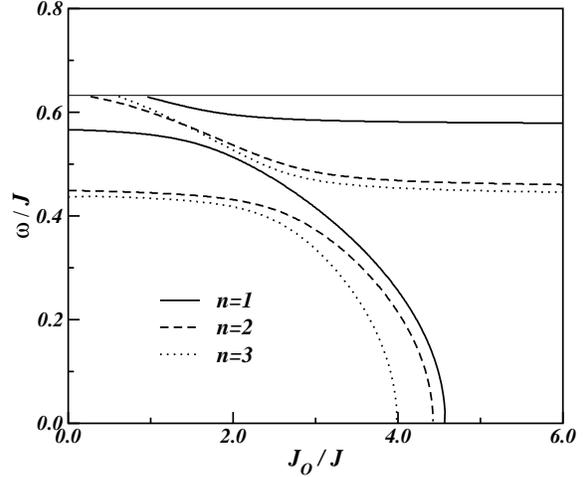}}
\caption{Localized spin wave frequencies as a function of the exchange
parameter $J_o$, for the $X$ case, with $h_o=h_{o'}=0.65h$ and
$T=2.5J/k_B$. The parameters are $J_{o'}=1.5J$ and $J_I=0.25J$ when
both impurities are at the surface (solid lines), and $J_{o'}=2.5J$,
$J_I=0.25J$ when the two impurities are at the second(dashed lines)
and third layers(dotted lines).}
\end{figure}

\section{Numerical Results}
Impurities modes spectra were calculated as functions of the exchange
and effective field parameters. In order to assess the influence of
the position of each impurity on the excitation spectra, we obtained
numerical solutions of Eq.(14) for impurities located at the surface
of the medium (layer 1) as well as deeper in the system (layers 2 and
3). Specifically, for the $X$ case, we considered three
configurations, namely, we set $n=1$ (both impurities at the surface),
$n=2$ (both impurities at the second  layer) and $n=3$ (both
impurities at the third layer) according to the  notation in Fig. 1a.
Likewise, for the $Z$ case, following Fig. 1b we set $n=1$ (i.e. one
of the impurities at the surface, the other at layer 2), $n=2$ (one at
layer 2, the other at layer 3) and $n=3$ (impurities at layers 3 and
4).

\begin{figure}[]
\resizebox{0.55\textwidth}{!}{\includegraphics{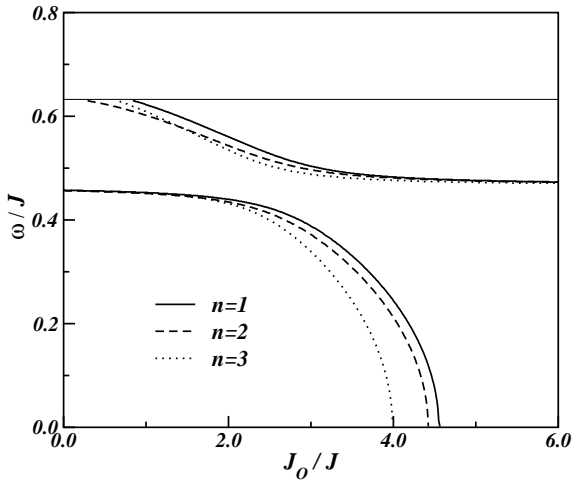}}
\caption{Localized spin wave frequencies as a function of the exchange
parameter $J_o$ in the {\it upper} impurity in an Ising ferromagnet
in a transverse field, for the $Z$ case, with $h_o=h_{o'}=0.65h$ and
$T=2.5J/k_B$. The parameters are $J_{o'}=1.5J$ and $J_I=0.25J$ when
the upper  impurity is located at the surface (solid lines),
and $J_{o'}=2.5J$, $J_I=0.25J$ when the upper impurity is at the
second (dashed lines) and third layers  (dotted lines).}
\end{figure}

Figure 2 shows the local SW impurities spectra for the $X$ case, as a
function of the exchange constant for the coupling between one of the
impurities and its neighboring host sites ($J_o$), while the
parameters for the second impurity are kept constant. We have used
$J_{o'}=1.5J$ and $J_I=0.25J$ for $n=1$, and $J_{o'}=2.5J$,
$J_I=0.25J$ for $n=2$ and $3$. The field parameters were
$h_o=h_{o'}=0.65h$ and we used $T=2.5J/k_B$ in all  cases.
The three sets of branches correspond to the impurities located
in layers 1 (solid line), 2 (dashed line) and 3 (dotted line). The
impurities modes can be classified as resonance modes (i.e. those
occurring in the SW bulk band) and defect modes (those found outside
the bulk  band). In this work we consider defect modes, since
these are easier to measure. Therefore we restrict the calculation to
the region outside the bulk SW band.

The lower limit of the bulk band is represented in the graph by the
horizontal line. In the absence of the coupling between the
impurities, the graph would show a horizontal line (for the impurity
with $J'$ constant) superimposed on the decaying curve of the SW
branch associated with the second impurity, which  merges with the
bulk band for low values of $J_o$. The exchange coupling thus creates
a mode repulsion effect when $J_o \approx J_{o'}$. As  expected,
the magnitude of this effect depends on the strength of the exchange
coupling between the impurities. When both impurities are at the
surface, the lower coordination number causes a large frequency shift
in comparison with the other configurations.

\begin{figure}[]
\resizebox{0.55\textwidth}{!}{\includegraphics{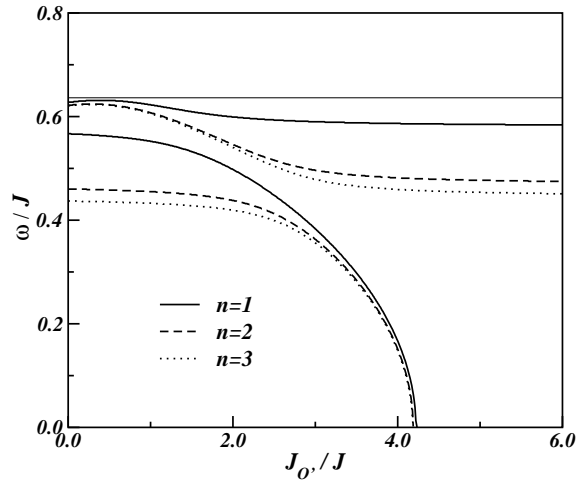}}
\caption{Localized spin wave frequencies as a function of the exchange
parameter $J_{o'}$ in the {\it lower} impurity in an Ising ferromagnet
in a transverse field, for the $Z$ case, with $h_o=h_{o'}=0.65h$ and
$T=2.5J/k_B$. The parameters are $J_{o}=1.5J$ and $J_I=0.25J$ when the
upper  impurity is located at the surface (solid lines),
and $J_{o}=2.5J$, $J_I=0.25J$ when the upper impurity is at the second
(dashed lines) and third layers  (dotted lines).}
\end{figure}

The effect of the alignment of the impurities is shown in Fig. 3,
which presents results for the SW frequencies in the $Z$ case,
as a function of the exchange constant at the {\it
upper} impurity, while the remaining parameters were kept constant.
For $n=2$ and $n=3$, the results are similar to the
ones obtained for $X$ case, apart from a small frequency shift. For
$n=1$, however, the branches are strongly shifted. These results
demonstrate that the influence of orientation becomes particularly
important for impurities close to the surface. This is a consequence
of the smaller number of neighbors, together with modified exchange
parameters at the surface.

Figure 4 shows results of frequency as a function of $J_{o'}$ at the
{\it lower} impurity in the $Z$ case, with the remaining parameters
kept constant. In this case, the exchange parameter being varied is
associated with the {\it lower} impurity (see Fig. 1), which was
assumed to be located at the second, third and fourth layers.
In contrast with the results of Fig. 3, the frequency branches
are not found to merge with the bulk band for low values of the
exchange parameter. Moreover, when the {\it lower} impurity is
localized at the second layer(solid line), the curves show a larger
shift in comparison with the remaining configurations. This effect is
a consequence of the modification of the exchange parameters for the
{\it upper} impurity when located at the  surface.

\begin{figure}
\resizebox{0.55\textwidth}{!}{\includegraphics{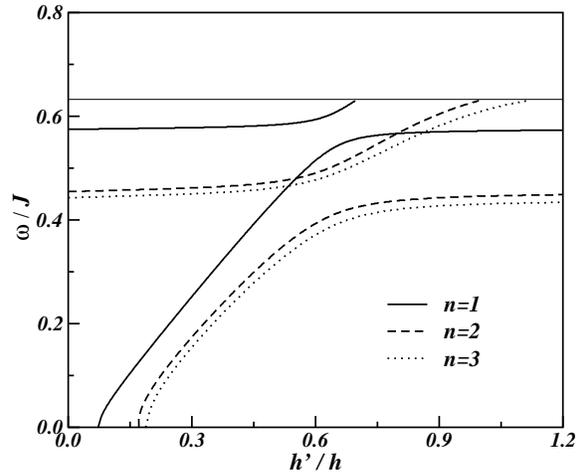}}
\caption{Localized spin wave frequencies as a function of the
local field parameter $h_o$ in one of the impurities, in an Ising
ferromagnet in a transverse field, for the $X$ case. The remaining
parameters are the same as in the previous graphs. The results were
obtained for both impurities at the first (solid lines), second
(dashed lines) and third layers (dotted lines).}
\end{figure}

The influence of the local fields is shown next. Figure 5 shows the
local SW frequencies as a function of local field parameter $h_o$
at one of the impurities, for the $X$ case. The remaining
parameters are the same as in the previous graphs. The results were
obtained for $n=1$ (solid lines), $n=2$ (dashed lines) and $n=3$
(dotted lines). The graphs show the mode repulsion, due to the
coupling of the impurities.  Also
evident is a large frequency shift for $n=1$, due to the modified
exchange parameters at the surface. For fields larger than
$h_o=0.69\,h$  ($n=1$), $h_o=0.99\,h$ ($n=2$) and $h_o=1.11\,h$
($n=3$), the {\it upper}  frequency branches are found to merge with
the bulk band, thus becoming  resonance
modes.

A similar graph is shown in Fig. 6, this time for the $Z$ case. The
results were calculated for a varying local field at the {\it upper}
impurity. In contrast with the previous figure, the higher frequency
branches in the graph do not display any noticeable frequency shift
for low values of the local field. On the other hand, the high
frequency  excitations are observed to become resonance modes for
values of local fields that approximately the same as in the $X$ case.

Figure 7 shows a plot of frequency as a function of the exchange
coupling constant for the interaction between the impurities ($J_I$).
The graph shows results for both the $X$ and $Z$ cases, and the
results were obtained  for ferromagnetic ($J_I>0$) as well as
antiferromagnetic ($J_I<0$)  coupling. For the $X$ case, we assumed
that both impurities were located at the  surface of the system,
whereas in the $Z$ case we considered the {\it upper} impurity at the
surface, and the {\it lower} one at the second layer. The branches
behave quite distinctly in each case, with the results for the $Z$
case displaying a large shift in comparison with the other case. Also
prominent in the $X$ case is a mode crossing that is found for
$J_I=-0.86J$. In contrast, the  results for the $Z$ case show no mode
crossings, although the low frequency is observed to have a maximum at
$J_I=-1.11J$ and the high frequency  branch has a minimum at
$J_I=-1.37J$.

\begin{figure}
\resizebox{0.55\textwidth}{!}{\includegraphics{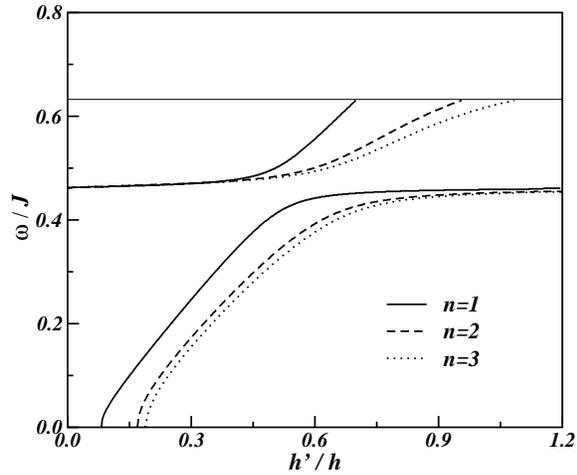}}
\caption{Localized spin wave frequencies as a function of the local
field parameter $h_o$ for the {\it upper} impurity, in an Ising
ferromagnet in a transverse field, for the $Z$ case, when the upper
impurity is located at the first (solid lines), second (dashed lines)
and third layers  (dotted lines).}
\end{figure}

\begin{center}
\begin{figure}
\resizebox{0.55\textwidth}{!}{\includegraphics{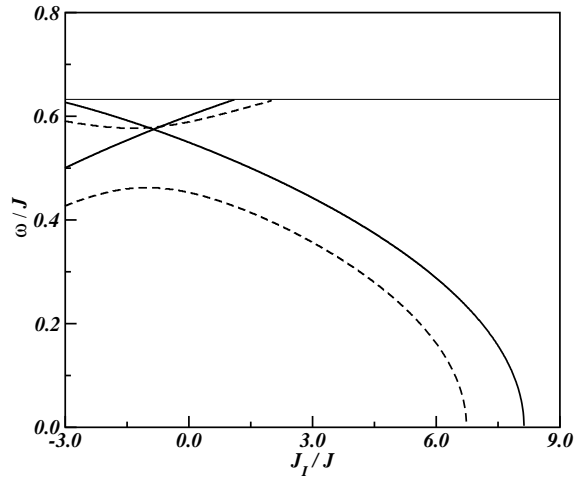}}

\caption{Localized spin wave frequencies as a function of the exchange
parameter $J_I$ for the coupling between the impurities. The graph
shows results for both the $X$ case (solid line) where the impurities 
are located at the surface layer, and the $Z$ case (dashed line) where
one impurity is located at the surface and the other in the layer
below. The remaining parameters are the same as in the previous
figures.} \end{figure} \end{center}

\section{Conclusions}
We have presented a Green's functions calculation of the SW
frequencies of localized modes associated with two coupled localized
impurities implanted in an  otherwise ideal
ferromagnet. The results were obtained in the context of the Ising
Model in a transverse field and show the influence of the exchange
coupling of the impurities on the excitation spectra of the system.
This coupling modifies the spectra  in relation to the results for
single impurities, especially when the coupling parameters  for the
interactions between the impurities and the host sites are of the same
magnitude. The results also point to a strong effect of the
orientation of the impurities on the localized SW  frequencies,
especially when the impurities are located close or at the surface. 
Further studies could investigate the effect of the
presence of the impurities on thin  ferromagnetic films, as well as
the effect of their coupling and orientation on the critical
properties of the system as well as on the overall magnetization of
the material.

The authors would like to acknowledge the financial support of the
Brazilian agency CNPq.

\end{document}